\documentclass{emulateapj}
\righthead{MU\~NOZ-DARIAS, CASARES \& MARTINEZ-PAIS}
\lefthead{K-CORRECTION IN LMXBs}
\begin{document}

\title{The "K-correction" for irradiated emission lines in LMXBs: evidence for a massive neutron star in X1822-371 (V691 CrA)}

\author{T. Mu\~noz-Darias,\altaffilmark{1 *} J. Casares,\altaffilmark{1} and I. G. Mart\'\i{}nez-Pais\altaffilmark{1,2}}

\altaffiltext{1}{Instituto de Astrof\'\i{}sica de Canarias, 38200 La Laguna,
Tenerife, Spain}
\altaffiltext{*}{tmd@iac.es}
\altaffiltext{2}{Departamento de Astrof\'\i{}sica, Universidad de La Laguna, E-38206 La Laguna, Tenerife, Spain}

\begin{abstract}

We study the K-correction for the case of emission lines formed in the X-ray illuminated atmosphere of a Roche lobe filling star. We compute the K-correction as function of the mass ratio $q$ and the disc flaring angle $\alpha$ using a compact binary code where the companion's Roche lobe is divided into $10^{5}$ resolution elements. We also study the effect of the inclination angle in the results. We apply our model to the case of the neutron star low-mass X-ray binary X1822-371 (V691 CrA), where a K-emission velocity  $K_{em}=300 \pm$8 km s$^{-1}$ has been measured by Casares et al. (2003). Our numerical results, combined with previous determination of system parameters, yields $1.61M_{\odot}\leq M_{NS} \leq 2.32M_{\odot}$ and $0.44M_{\odot}\leq M_{2} \leq 0.56 M_{\odot}$ for the two binary components(i. e. $0.24 \leq $q$ \leq 0.27$), which provide a compelling evidence for a massive neutron star in this system. We also discuss the implications of these masses into the evolutionary history of the binary.

\end{abstract}


\section{Introduction}
Low mass X-ray binaries (LMXBs) are interacting binaries where a low
mass star transfers matter onto a neutron star (NS) or black hole (e.g. Charles
\& Coe 2004). There are $\sim$200 bright (L$_{X} \simeq 10^{36}-10^{38}$ erg
s$^{-1}$) LMXBs in the Galaxy and most of them harbour neutron stars as
implied by the detection of X-ray bursts/pulsations. About
$\sim$20 of these have period determinations but only in very few cases (e.g. Cyg X-2:
Casares et al. 1998, Orosz \& Kuulkers 1999 and V395 Car: Shahbaz et al. 2004, Jonker et al. 2005), a full orbital solution
and mass determination exists thanks to the spectroscopic detection of the giant companion star.
For the others, the companion is completely swamped by reprocessed light from the accretion flow
and this has hampered detailed knowledge of their dynamical properties.\\

X1822-371 (V691 CrA) is one of the brightest LMXBs in the optical,
and also the prototypical accretion disc corona (ADC) source due to its low $L_x/L_{opt}$ $\sim$20 ratio (it is usually $\sim$500-1000). In ADC sources, the combination of high inclination and a thick disc obscures the central source, and only X-rays scattered from material above and below the disc can reach the observer (White \& Holt 1982). Thanks to eclipses of the accretion disc by the companion star on a 5.57h orbital period (Hellier \& Mason 1989; hereafter HM89)
an accurate determination of the inclination angle exists, i=82.5 $\pm$1.5 (Heinz \& Nowak 2001; hereafter HN01). The source also displays 0.59s X-ray pulses (Jonker \& Van der Klis 2001) which  enables an extremely precise determination of the neutron star radial velocity semiamplitude of $K_1$=94.5$\pm$0.5 km s$^{-1}$. Therefore, the only parameter missing for an accurate mass determination in X1822-371 is the radial velocity curve of the companion star.

HeI absorption features attributed to the irradiated hemisphere of the companion star were presented in Harlaftis, Horne \& Charles (1997), Jonker et al. (2003) and Cowley et al. (2003). They measured  $K_{2}>225\pm23$ km s$^{-1}$, $>280\pm26$ km s$^{-1}$ and $>234\pm20$ km s$^{-1}$ respectively  by fitting sine waves to the HeI radial velocity curves. However, as Jonker et al. pointed out the spectroscopic T$_0$ is delayed by $0.1P_{orb}$ with respect to the photometric eclipses and pulsar ephemeris and hence asymmetric irradiation of the Roche lobe needs to be invoked.

In Casares et al. (2003; hereafter C03), the first unambiguous detection of the donor star is reported using the NIII $\lambda$4640-Bowen emission line with a velocity amplitude of $K_{em}=300\pm8$ km s$^{-1}$.
The phasing of the NIII velocity curve is perfectly consistent with the pulsar ephemeris with no need for ad hoc asymmetric illumination effects. Furthermore, they show that the HeI absorptions come from the accretion stream and not from the companion star. This naturally explains the phase offset with the pulsar ephemeris observed in previous works.

However, the NIII-Bowen emission line is excited on the inner hemisphere of the donor star and only provides a lower limit to the true $K_{2}$-velocity.
Narrow high-excitation emission lines originating from the companion star have also been detected in the prototypical LMXB Sco X-1 (Steeghs \& Casares 2002), and in the black hole X-ray transient GX339-4 during its 2002 outburst (Hynes et al. 2003). These narrow features are strongest in the Bowen blend,
a set of resonant emission lines (mainly CIII and NIII $\lambda 4640$) excited by photoionization and the Bowen mechanism (McClintock, Canizares \& Tarter 1975). The Roche lobe shaped donor star intercepts the energetic photons from the inner accretion disc resulting on the observed optical emission lines
from its surface. The Doppler motion of these lines trace the orbit of the
companion star and provide the first reliable binary parameters in persistent LMXBs.
However, the secondary is not uniformly illuminated by the ionizing photons and, therefore, the radial velocity curve of the sharp features traces the motion of the reprocessed light-center rather than the true center of mass of the star.\\

In this paper we model the deviation between the reprocessed light-center and the center of mass of a Roche lobe filling star (the so-called "K-correction") in a persistent LMXB, including screening effects by a flared accretion disc. We apply our simulations to the case of X1822-371 and conclude that it must contain a massive NS with $M_{NS} \sim 1.61M_{\odot}-2.32M_{\odot}$, for any possible value of the disc flaring angle.

\section{The "K-correction" for (emission) lines in the irradiated face}

K-correction for radial velocity curves derived from absorption lines formed on the companion's photosphere was presented by Wade \& Horne (1988; hereafter WH88). In this case, absorption lines are quenched in the heated face and hence the observed K-velocity provides only an upper limit to the real $K_2$, i.e. the true velocity of the center of mass of the companion star. Here, we study the K-correction for the case of irradiated-induced spectral lines originating from the heated face of the donor star, which is the opposite case to the one presented in WH88. This problem was firstly addressed by Horne \& Schneider(1989; hereafter HS89) but it was only applied to the particular case of nova V1500 Cyg without screening effects by a possible accretion disc. Also Beuermann \& Thomas (1990) studied this problem for the case of the cataclysmic variable IX Velorum. They considered only two different accretion disc angles together with the orbital parameters of this binary. 
In this paper we tackle the problem in a general approach and derive a K-correction expression for all possible $q$ values and obscuring disc elevations. 

Following the prescription of HS89, the light-center of the emission line region($lc)$ is given by $lc=a_{2}-d$, where $d$ represents the light-center displacement from the center of mass of the donor star and $a_{2}=a/(1+q)$ is the distance between the center of mass of the companion and the center of mass of the system. $q= M_{2}/M_{1}$ is the mass ratio and $a$ the binary separation. Using this nomenclature, the difference between the $K_{2}$ and the observed emission line velocity ($K_{em}$) is

\begin{center}
$\Delta K=K_{2}-K_{em}=K_{2} \frac{d}{a_{2}}$
\end{center}

and therefore,

\begin{equation}
K_{2}=\frac{K_{em}}{1-d\frac{(1+q)}{a}} =\frac{K_{em}}{1-f(1+q)}
\end{equation}

where $f$ is an dimensionless factor which represents the  displacement of the emission light-center from the center of mass of the companion in units of $a$. In the limit case where all the emission line is formed at the inner Lagrange point($L_{1}$)

\begin{center}
$f = \frac{b_{2}}{a} \sim 0.5+0.227\log q$	(Plavec \& Kratochvil 1964)
\end{center}

where $b_{2}$ is the distance between $L_{1}$ and the center of mass of the companion star. On the other hand, we can consider the extreme case where the emission line comes from the irradiated point with maximum radial velocity. This point is located at the limb of the irradiated region and holds an opening angle $\alpha_{M}$ above the plane of the binary \footnote{Note that for this case to work the accretion disc must screen the rest of the inner face of the donor}. This is the highest possible angle for the companion to be irradiated and it has been tabulated by R. Pennington in Table A3 of the Appendix to Pringle \& Wade (1985). An approximate analytical expression, as a function of q, is provided by Paczynski equation (1971),

\begin{equation}
\sin{\alpha_{M}} \cong \frac{R_{2}}{a} \cong 0.462(\frac{q}{1+q})^{1/3} .
\end{equation}

Then, if we assume that the Roche lobe is spherical (which is a good approximation when we are far from $L_{1}$) it is easy to show that the minimum $f$ displacement is given by $\sim \sin^2{\alpha_{M}}$ . Therefore, in any possible situation  $f$ will always be constrained between these two cases i.e.

\begin{center}
$0.5+0.227 \log q>f>\sin^2{\alpha_{M}}$
\end{center}

These limits are presented in Fig. \ref{figf}, plotted in dashed line.
Here we have used the exact $\alpha_{M}$ values calculated by our code (see below) which obviously  coincide with the ones tabulated by R. Pepringter.  

\begin{figure}
\plotone{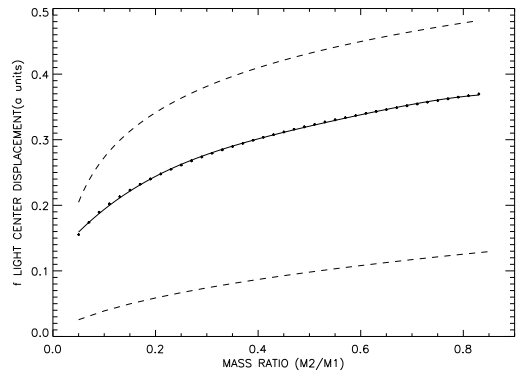}
\caption{$f$ factor or displacement of the emission light-center from the center of mass of the companion, as a function of $q$. Dashed lines represent the limit cases corresponding to $f \sim \sin^2{\alpha_{M}}$ (lower curve) and $f \sim \frac{b_2}{a}$ (upper curve). A more realistic upper limit to $f$, derived from the irradiation model is represented by a dotted line with the best polynomial fit overplotted in solid line as well.
\label{figf}}
\end{figure}

However, the upper limit obtained is purely theoretical since $L_{1}$ will never be the only irradiated point of the donor star. A more realistic upper limit can be obtained by simulating emission lines formed by isotropic irradiation on the inner hemisphere of the companion in a Roche lobe geometry. We have divided the Roche lobe filling star in $10^{5}$ triangular tiles of equal surface. The flux from the X-ray source which is reprocessed in each tile will be given by, 

\begin{center}
$F_{r} \sim \frac{F_{X}(1-\varepsilon) \cos{\theta_{X}}}{r^{2}}a_t$
\end{center}

where $F_{X}$ is the X-ray flux emitted by the compact object, $a_t$ the area of the tile, $r$ the distance between the tile and the center of mass of the primary star, $\cos{\theta_{X}}$ the projection factor of the tile as seen by the X-ray source and $\varepsilon$ the albedo. The total reprocessed emission line flux seen by a distant observer is obtained by integrating $F_{r} \cos{\theta_{v}}P_g$ over the Roche lobe geometry, where $\cos{\theta_{v}}$ is the projection factor of each irradiated tile as seen by the observer and $P_g$ represent a local gaussian profile. We have not included limb-darkening effects in the simulation because these are expected to be negligible in the optically thin case that we are considering here. To test this hypothesis, we have also included in our simulations a linearized limb darkening, appropriate to the continuum (i.e. optically thick case), and found it had no impact on the results.\\
\begin{figure}
\plotone{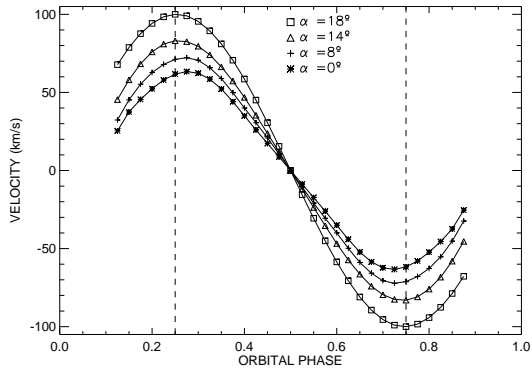}
\caption{Radial velocity curves for $q=0.6$, $i=90^{\circ}$ and four different disc obscuring angles $\alpha$. The plot shows how the curves start to deviate from sinewaves for low $\alpha$ values. The amplitude of the velocity curves depends on $\alpha$ as well. Note that the velocity values have been rescaled by normalizing to $v=100$ km s$^{-1}$ the maximum of the radial velocity curve corresponding to $\alpha=18^{\circ}$.
\label{figc}}
\end{figure}

\begin{figure}
\plotone{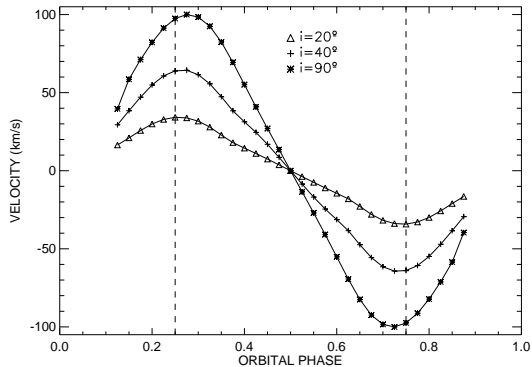}
\caption{Radial velocity curves for $q=0.6$, $\alpha=0^{\circ}$ and three different values of the inclination angle. The maxima of the radial velocity curves are closer to orbital phase 0.25 for the lower inclination cases. Note that the velocity values have been rescaled by normalizing to $v=100$ km s$^{-1}$ the maximum of the radial velocity curve corresponding to $i=90^{\circ}$.
\label{figci}}
\end{figure}

Radial velocity curves were subsequently computed by measuring the position of the maxima in the synthetic profiles (which are not symmetric), and $K_{em}$ is determined from the maxima of the radial velocity curves.
Therefore, for a given $K_{2}$ eq (1) gives the $f$ values as a function of $q$. These values, as expected, are not found to depend on either $\varepsilon$ nor $F_{X}$. The synthetic $f(q)$ curve is plotted in Fig. \ref{figf} in solid line and a 4th order polynomial fit (which is  accurate to better than 1\%) to this curve yields
\begin{center}
$f(\alpha=0^{\circ})=0.116 +  0.946 $q$ - 1.947 q^2 + 2.245 q^3  -1.002 q^4$
\end{center}
Then, the real value of $f$ will be constrained between these two limits i.e. $f(\alpha=0^{\circ})>f>\sin^2{\alpha_{M}}$ and will depend on the size of the companion (i.e. the mass ratio) and the height of the accretion disc rim obscuring the heated star.\\

\begin{figure}
\plotone{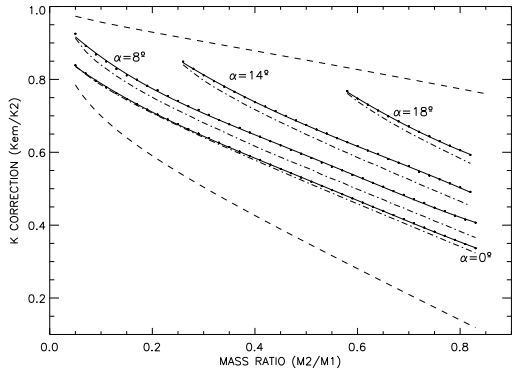}
\caption{K-correction as function of $q$ for different disc opening angles. Dotted lines represent the K-correction computed by our model for $i=90^{\circ}$ and disc opening angles $0^{\circ}$, $8^{\circ}$, $14^{\circ}$ and $18^{\circ}$. The solid lines mark the polynomial fits to the computed curves and the dash-dotted lines show the case of  $i=40^{\circ}$. The dashed lines represents the same limits as in figure 1.
\label{figkc}}
\end{figure}

In a second step, we have incorporated shadowing effects by a flared and opaque accretion disc with an opening angle $\alpha$ independent of azimuth.
Figure \ref{figc} presents the synthetic radial velocity curves computed for four different disc obscuring angles. The plot shows
how the radial velocity curves start to deviate from pure sinewave
functions when $\alpha$ decreases.  Because the velocity amplitude is strongly dependent on $\alpha$ so it  does the K-correction.
Therefore, using our binary code we have computed the relation between the ratio $\frac{K_{em}}{K_{2}}$ and $q$ for different values of the flaring angle $\alpha$ and these are plotted in Fig. \ref{figkc} in solid line.\\
In Fig. \ref{figci} we also explore the effect of the inclination angle in the shape of the radial velocity curve. This plot shows that the morphology of the radial velocity curves clearly depends on the inclination angle and hence, $\frac{K_{em}}{K_2}$ could be a function of this parameter as well. Using our numerical model we have found that (in the most extreme cases when $q> 0.6$), $\frac{K_{em}}{K_2}$ can be $\sim 10\%$ smaller for low inclination angles than for  $i=90^{\circ}$. To illustrate the effect of the inclination angle we have also computed the K-correction for $i=40^{\circ}$ and different $\alpha$ values, and these are  overplotted in Fig. \ref{figkc} in dash-dotted line.\\
The synthetic K-corrections produced by the numerical code are very smooth and can  be well described by fourth-order polynomial fits. Table 1, lists the coefficients of the fits for the two values of the inclination considered and different opening angles (from $0^{\circ}$ to $18^{\circ}$) which cover the typical values observed in XRBs and cataclysmic variables. These fits are accurate to better than 1\% between $q \sim 0.05$ and $q \sim 0.83$. For the case of  intermediate inclinations the interpolated solution  between $i=90^{\circ}$ and $i=40^{\circ}$ is accurate to $\leq2$\%. For $i\leq40^{\circ}$ the $i=40$ value gives also a very good aproximation ($< 2$\%). Note that the value of $\frac{K_{em}}{K_2}$ obtained for $i=90^{\circ}$ yields a secure lower limit to the masses of the binary components for all possible cases.\\
For intermediate $\alpha$ and $i$ values one needs to interpolate the K-correction solutions and not the coefficients.

\placetable{tab1}

Hence, for a given  mass ratio $q$, the K-correction is provided by:

\begin{equation}
\frac{K_{em}}{K_{2}} \cong N_0 + N_1 q + N_2 q^2 + N_3 q^3 + N_4 q^4
\end{equation}

Note that, since the K-correction is only meaningful if $\alpha \leq \alpha_{M}$ (otherwise the companion will be completely shadowed by the disc), eq. (2) sets a lower limit to q for each $\alpha$, (see Fig 4.). For this reason the first values of $q$ considered in the fits for $\alpha$= $10^{\circ}$, $12^{\circ}$, $14^{\circ}$, $16^{\circ}$ and $18^{\circ}$ are $q$= 0.1, 0.15, 0.26, 0.4 and 0.58 respectively. As a test, we have computed the K-correction for the binary parameters of nova V1500 Cyg and obtain $\frac{K_{em}}{K_{2}} = 0.64$ in excellent agreement with HS89.
\section{The "K-Correction" for X1822-371}
\begin{figure}
\plotone{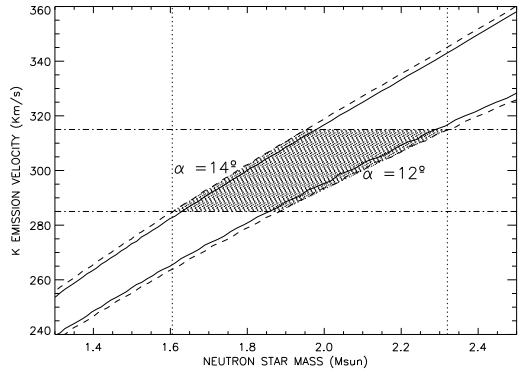}
\caption{Determination of the neutron star mass in X1822-371. The solid lines represent the results of the model for the opening angles considered ($\alpha \sim 12^{\circ}$ and $\alpha \sim 14^{\circ}$). The dashed lines represent the uncertainty in the orbital inclination. We have also marked in dot-dashed lines the measured $K_{em}=300 \pm 15 km s^{-1}$. The possible neutron star mass range is restricted by the shadowed region.
\label{fig4}}
\end{figure}

Now we can apply our computed K-correction to the case of X1822-371 and derive the masses for the two components. In C03
a $K_{em}$= 300 $\pm$8 km s$^{-1}$ was reported through the detection of the NIII $\lambda 4640$-Bowen emission line arising from
the companion star. This value, combined with the previous determination of the inclination angle and the pulsar's
radial velocity curve, yields absolute lower limits $M_{1} \geq 1.14 M_{\odot}$ and $M_{2} \geq 0.36 M_{\odot}$, assuming $K_{2} \geq K_{em}$ (see C03).\\
On the other hand, the mass function equation can be expressed as
$$M_1 = \frac{(\frac{K_1}{q})^3 P(1+q)^2}{2 \pi G \sin^3{i}}$$
where P is the orbital period and G is the gravitational constant. In the case of X1822-371, where $i$ and $K_{1}$ are known, this equation provides a direct relation between $q$ and $M_{1}$. Therefore, this can be combined with our synthetic K-correction curves to yield $K_{em}$-$M_{1}$ relations for different disc opening angles $\alpha$.\\
C03 obtained $K_{em}= 300 \pm 8$ km s$^{-1}$ by fitting the NIII$\lambda 4640$ spot  in the Doppler map with a 2D-gaussian function. Instead, we have computed the centroid of the spot intensity distribution  and obtain a more conservative value of $K_{em}$= 300 $\pm$15 km s$^{-1}$, which we prefer since it does not impose any model to the fit. In order to test whether the K-correction derived by our model can be used to correct the $K_{em}$  value extracted from the Doppler tomography we have
simulated irradiated profiles (using our derived  parameters of X1822-371, see below)
for the orbital phases of the C03 observations. These were smeared according to the exposure times used and subsequently degraded to the  spectral resolution of the observations. Finally, we added noise to reproduce the observed spectra. A Doppler map was then computed and the centroid of the spot was found to be in excellent agreement, within the errors, with the value above. Furthermore, we note than the same $K_{em}$  was obtained by SC02 in Sco X-1 from the radial velocity curve and Doppler tomography.\\
Under the assumption that the Bowen emission lines are formed on the critical Roche lobe of the companion the detection of this emission at $K_{em}= 300 \pm$ 15 km s$^{-1}$ directly yields $M_{NS} \gg 1.35M_{\odot}$ and  $M_{2} \gg 0.4M_{\odot}$ through the geometrical limit $f>\sin^2{\alpha_{M}}$. However, realistic mass determinations require the knowledge of the disc rim elevation obscuring the companion. De Jong et al. (1996) derived  a mean disc opening angle of $\alpha \sim 12^{\circ}$  by studying the effects of X-ray reprocessing in LMXBs. This value was also derived by HM89 for the highest disc structure (i.e. the bulge at phase $\sim$0.8) in X1822-371 by modeling simultaneous optical and X-ray EXOSAT lightcurves. However, in HN01 a higher angle of $\alpha \sim 14^{\circ}$ is obtained by fitting X-ray RXTE+ASCA lightcurves. These estimates correspond to the disc elevation at the phase of the bulge, and therefore, must be treated as upper limits to the height of the disc rim obscuring the companion, which is constrained to $-0.04 \leq \phi \leq 0.04$. In any case, these opening angles  are expected to be, in general, only slightly higher than the real ones, and hence, we have calculated the $K_{em}$-$M_1$ relation for both disc thicknesses as it is shown in Fig. \ref{fig4}. If we take into account, the opening angles considered, the inclination uncertainty and the errors in the observed $K_{em}$, Fig. \ref{fig4} shows that the mass of the neutron star is constrained to the range
\begin{center}
$1.61M_{\odot}\leq M_{NS} \leq 2.32M_{\odot}$.
\end{center}
We would like to remark that this provides a conservative lower limit to $M_{NS}$. If the accretion disc were thinner the mass would be even higher. For instance, if  $\alpha = 0^{\circ}$ (i.e. no disc), $M_{NS}$ must be $\simeq$ $3M_{\odot}$ in order to reproduce the observed $K_{em}$ at $\simeq 300$ km  s$^{-1}$. Besides, since the neutron star mass and the mass ratio are related by the mass function we obtain $0.24 \leq $q$ \leq 0.27$ which leads to
\begin{center}
$0.44M_{\odot}\leq M_{2} \leq 0.56 M_{\odot}$.
\end{center}
\section{discussion}
The detection of irradiated-induced emission lines originating from the heated face of the companion star can yield important restrictions to the mass of the binary components.  $K_{em}$ contains intrinsic information about the companion's regions responsible for reprocessing the X-ray/UV radiation. Since the orbital parameters of X1822-371 are tightly constrained, the key missing information for a realistic simulation are the structures present between the companion and the X-ray source. Our simple model only takes into account screening effects by a flared accretion disc. However, the presence of a disc corona scattering the X-rays/EUV radiation might modify the results of our simulations since irradiating photons could reach (otherwise obscured) regions of the Roche lobe with lower velocities. Consequently, the scattered radiation is expected to push the K-correction to lower values and hence the NS to higher  masses.
As a test we have included emission from a 'corona' which  has been  roughly simulated by five  X-ray spots symmetrically distributed around the compact object at  a distance of $0.2a$ (i.e. the 'corona' radii derived in HN01). The integrated flux of this simplified 'corona' was scaled to be 1\% of the accretion luminosity i.e. 10$^{36}$ erg s$^{-1}$ for an assumed compact object luminosity of  10$^{38}$ erg s${-1}$. Our simulation shows that the effect of such a 'corona' in the K-correction is completely negligible. Even if we rise the luminosity of the corona to 10\% of the central X-ray source and take $\alpha > 10^{\circ}$, the effect in $K_{em}$ is lower than 1\%. In this extreme case, the mass of the NS would only rise by $\sim 0.03$ M$_{\odot}$ with respect to neglecting the effect of corona scattering.
In summary, our simulation provides compelling arguments for the presence of a massive NS in X1822-371\\
To date, the most accurate neutron star mass determinations have been provided by studying radio pulsars. These works reflect
that the mass of the neutron star at formation is $\sim 1.35 \pm 0.04 M_{\odot}$ (van Kerkwijk 2001). However, recent studies
have revealed evidences for heavier neutron stars in old white dwarf + pulsar systems, especially in J0751+1807 where
a mass between 1.6-2.8 $M_{\odot}$ has been reported by Nice, Splaver, \& Stairs (2004). There is
no compelling evidence for massive neutron stars in active XRBs yet, but possible candidates
are Cyg X-2, with $M_{NS} =1.78 \pm 0.23 M_{\odot}$ (Casares et al. 1998; Orosz \& Kuulkers 1999)
Vel X-1, with $M_{NS} = 1.86 \pm 0.33 M_{\odot}$ (Barziv et al. 2001) and V395 Car, with $M_{X} =1.9-4.1 M_{\odot}$ (Shahbaz et al. 2004; Jonker et al. 2005). In the latter case the nature of the compact object is an open issue since NS signatures (X-ray burst, pulses) have not yet been detected. The evidence for a massive NS in X1822-371 reinforces the theory that LMXBs could harbour heavy neutron stars due to sustained episodes of high mass transfer.
\subsection{The nature of the companion star and evolutionary history of X1822-371}

Faulkner, Flannery \& Warner (1972) showed that the mean density of a Roche lobe-filling companion star is determined solely by the binary period $P$:

\begin{center}
$\bar{\rho} \cong 113P_{hr}^{-2}$ g cm$^{-3}$
\end{center}

In the case of X1822-371, with $P_{hr}=5.57$, we obtain $\bar{\rho} \cong 3.7$ g cm$^{-3}$. This value is consistent with a $0.5M_{\odot}$ M0 main sequence(MS) star (Cox 2000) and it is in excellent agreement with our derived mass. We also note that a $0.5M_{\odot}$ MS star would not have evolved in a Hubble time. However, if we assume that the NS at formation was $\sim 1.4M_{\odot}$ it must have accreted $\sim 0.5M_{\odot}$ from the companion to reach its current $\sim 1.9M_{\odot}$ and hence,  it must have had at least $\geq 1M_{\odot}$ at the onset of the mass transfer phase.\\
Parmar et al. (2000; see also Baptista et al 2002) demonstrated that the orbital period of X1822-371 is increasing at $\dot{P} = 1.78$ x $10^{-10}$. As it was discussed in HN01, this period variation can be used to determine the  mass transfer rate from the companion $\dot{M_{2}}$, which, in a conservative case, equals the  accretion rate onto the NS ($\dot{M_{1}}$). The $\dot{M_{1}}$ value implied by our $\sim1.9M_{\odot}$ NS is $\sim 3-6$ x $10^{-8} M_{\odot}$yr$^{-1}$ ( comparable to the one obtained for an assumed $1.4M_{\odot}$ NS) and yields $L_{X} \sim  10^{38} $erg s$^{-1}$. Since X1822-371  is an ADC source we only see X-rays  scattered by the "corona" and there are only rough estimates of the true luminosity. For instance, Jonker \& Van der Klis (2001) derived $L_{X} \sim 2-4$ x $10^{37}$ erg s$^{-1}$ from the spin-up and constraints to the magnetic field. This, in turn,  imply $\dot{M_{1}} \sim 10^{-9} M_{\odot}$yr$^{-1}$ i.e. one order of magnitude lower than predicted from $\dot{P}$. However, this discrepancy could be reconciled if $\dot{M_{1}}$ were driven by secular changes and not reflect $\dot{P}$ (e.g. magnetic cycles in the companion ;Hellier et al. 1990).\\
HN01 proposed that X1822-371 might have had a evolutionary history similar to Cyg X-2 (Podsialowski \& Rappaport 2000 ; Kolb et al. 2000) where the system descend from  an intermediate mass X-ray binary (IMXB) with $q>1$. In this scenario the secondary  must have lost more than $\sim 1M_{\odot}$ from its outer layers in order to reach its current mass, and it should expose CNO processed material from the inner layers. Unfortunately, the companion star is not directly detected spectroscopically and hence the spectral type  and the history evolution are unconstrained. However, the emission line spectra reported in C03  seem to show "normal-strength" CIII lines within the Bowen blend  and also a strong OVI at $\lambda3811$. Our observed masses, orbital period and $\dot{M_{1}}$ could be accommodated by some IMXB evolutionary sequences with an initial companion's mass and orbital period  of $\sim 3.5M_{\odot}$ and $\sim20$hr respectively (Ph. Podsialowski private communication). The currently observed $q\sim 0.25$ implies that X1822-371 must have long passed the thermal mass transfer phase and its evolution should be currently driven by magnetic braking phase with $\dot{P}<0$ to sustain mass transfer, in clear contradiction with the observed $\dot{P}>0$. Alternatively, mass transfer could be sustained by X-ray irradiation as it was suggested in Pfahl, Rappaport \& Podsialowski (2003). A fundamental test of the evolutionary history of X1822 will be to perform new high resolution optical and UV spectroscopy to search for evidence of CNO processed material (eg. XTE J1118+480, Haswell et al. 2002).\\

We thank Ph. Podsialowski for his useful comments. We are also grateful to the referee for his useful comments and suggestions which have improved an earlier version of the manuscript. JC acknowledges support by the Spanish MCYT grant AYA2002-0036.

\begin{table}
\begin{center}
\caption{K-correction fit coefficients
\label{tab1}}
\begin{tabular}{c | c c c c c | c c c c c}
~ & \multicolumn{5}{|c}{$i=90^{\circ}$} & \multicolumn{5}{|c}{$i=40^{\circ}$} \\
\hline
$\alpha$ & $N_{0}$ & $N_{1}$ & $N_{2}$ & $N_{3}$ & $N_{4}$     &    $N_{0}$ & $N_{1}$ & $N_{2}$ & $N_{3}$ & $N_{4}$\\
\hline
$0^{\circ}$ & 0.888 & -1.291 & 1.541 & -1.895 & 0.861 	       &    0.886 & -1.132 & 1.523 & -1.892 & 0.867 \\
$2^{\circ}$ & 0.903 & -1.151 & 1.591 & -1.914 & 0.844          &    0.897 & -1.159 & 1.591 & -1.935 & 0.860\\
$4^{\circ}$ & 0.922 & -1.213 & 1.855 & -2.358 & 1.106          &    0.904 & -1.477 & 1.576 & -1.948 & 0.888\\
$6^{\circ}$ & 0.942 & -1.230 & 1.877 & -2.336 & 1.070 	       &    0.929 & -1.313 & 2.147 & -2.733 & 1.264\\
$8^{\circ}$ & 0.982 & -1.430 & 2.552 & -3.284 & 1.542          &    0.975 & -1.559 & 2.828 & -3.576 & 1.651\\
$10^{\circ}$ & 1.015 & -1.444 & 2.426 & -2.957 & 1.333         &    1.015 & -1.590 & 2.696 & -3.233 & 1.439\\
$12^{\circ}$ & 1.093 & -1.901 & 3.593 & -4.280 & 1.888         &    1.130 & -2.160 & 4.115 & -4.786 & 2.062\\
$14^{\circ}$ & 1.137 & -1.418 & 1.324 & -0.679 & 0.000         &    1.165 & -1.617 & 1.565 & -0.797 & 0.000 \\
$16^{\circ}$ & 1.326 & -1.871 & 1.745 & -0.782 & 0.000         &    1.428 & -2.362 & 2.421 & -1.122 & 0.000 \\
$18^{\circ}$ & 1.496 & -1.645 & 0.665 & 0.000 & 0.000           &    1.579 & -1.859 & 0.767 & 0.000 & 0.000 \\
\hline
\end{tabular}
\end{center}
\end{table}
\end{document}